\documentclass[preprint,tightenlines,nofootinbib,floats,tighten,prd,aps]{revtex4}
\usepackage{amsmath}
\usepackage{epsf}

\def\be{\begin{equation}}
\def\ee{\end{equation}}
\def\half{{1\over 2}}
\def\bea{\begin{eqnarray}}
\def\eea{\end{eqnarray}}
\def\bml{\begin{subequations}}
\def\blea{\begin{subequations}\begin{eqnarray}}
\def\elea{\end{eqnarray}\end{subequations}}
\def\diag{\mathop{\rm diag}}

\def\ba{{\bf a}}
\def\bb{{\bf b}}
\def\bx{{\bf x}}
\def\bbeta{{\bbox{\beta}}}
\def\balpha{{\bbox{\alpha}}}
\def\betapar{{\beta_\parallel}}
\def\betaperp{{\bbeta_\perp}}
\def\betahatperp{{\widehat\bbeta_\perp}}
\def\xdot{\dot{\bf x}}
\def\atilde{{\tilde{\bf a}}}
\def\btilde{{\tilde{\bf b}}}
\def\ttilde{{\tilde t}}
\def\xtilde{{\tilde {\bf x}}}
\def\Atilde{{\tilde A}}
\def\Btilde{{\tilde B}}
\def\mutilde{{\tilde\mu}}

\begin{document}
\draft
\title{The form of cosmic string cusps}

\author{J.\ J.\ Blanco-Pillado\footnote{Email address: {\tt
jose@cosmos2.phy.tufts.edu}} 
and Ken D.\ Olum\footnote{Email address: {\tt kdo@alum.mit.edu}}}

\address{Institute of Cosmology \\
Department of Physics and Astronomy \\
Tufts University \\
Medford, MA 02155}

\date{October 1998, corrected January 2021}

\begin{abstract}
We classify the possible shapes of cosmic string cusps and how they
transform under Lorentz boosts.  A generic cusp can be brought into a
form in which the motion of the cusp tip lies in the plane of the
cusp.  The cusp whose motion is perpendicular to this plane,
considered by some authors, is a special case and not the generic
situation.

We redo the calculation of the energy in the region where the string
overlaps itself near a cusp, which is the maximum energy that can be
released in radiation.  We take into account the motion of a generic
cusp and the resulting Lorentz contraction of the string core.  The
result is that the energy scales as $\sqrt {rL}$ instead of the usual
$r ^ {1/3} L ^ {2/3}$, where $r$ is the string radius and $L$
is the typical length scale of the string.  Since $r\ll L $ for
cosmological strings, the radiation is strongly suppressed and could not
be observed.

\end{abstract}


\maketitle

\section{Introduction}

Cosmic strings are topologically stable defects which may have
formed during a symmetry breaking phase transition in the early
universe.  Due to their potential cosmological and astrophysical
effects, their properties and evolution have been extensively studied
in the past decades.  (For reviews see \cite{Alexbook,Kibble95}). 
 After formation, strings evolve into a scaling
network in which a Hubble volume at any time contains of order one
long string and a larger number of oscillating loops in the process of
decay.  The decay of cosmic string loops has been studied as a
possible source of ultra-high-energy cosmic rays\cite{Alex-Venia98,Pijus98}.

When the string is smooth on scales of the order of the string
thickness, the equations of motion reduce to the Nambu-Goto
form, in which the string is treated as a one-dimensional object
without thickness.  A typical cosmic string loop, in the course of its
oscillations, produces one or more cusps: points where, in the
Nambu-Goto approximation, the string attains the velocity
of light, and doubles back on itself.  Near a cusp, the
string can interact with itself and some of the energy stored in the
string can be released as high-energy radiation
\cite{Branden87,Pijus89,Branden90,Branden93a,Branden93b}.  In this paper we
analyze the possible forms of such cusps and discuss how the shape
and motion are affected by the choice of Lorentz frame.

We will consider a gauge string with energy scale $\eta$, which leads
to a string tension $\mu \sim \eta^2$ and a core radius
$r\sim\eta^{-1}$.  The first deviation from the Nambu-Goto equations
of motion appears at order $(r/ R)^4$ \cite{Anderson97} where $R$ is
the radius of curvature of the string.  Far from any cusp (or kink),
$R$ is of cosmological size, whereas $r$ is tiny, so the Nambu-Goto
approximation is very good.  Near the cusp, the two branches of string
can interact with each other, and the Nambu-Goto approximation is no
longer accurate.  However, any such interaction must fall off rapidly
at large separations, so string points whose closest distance of
approach is several times $r$ will not be affected by any corrections,
and thus will have no chance to interact.  Thus the result of the
corrections can be, at most, to change the effective radius at which
the two branches of the string can be considered to overlap.

Vincent, Antunes and Hindmarsh \cite{hind98} found significant
departures from Nambu-Goto evolution in a field theory simulation of a
string with standing waves, but Moore and Shellard \cite{Shellard98}
did not find such an effect.  We have simulated cusp production in
lattice field theory \cite{JJKDO98.1} and have not found a significant
departure from the Nambu-Goto evolution before the time of the cusp.
We will assume here that the amount of energy emission can be
calculated by following the Nambu-Goto equations of motion and finding
those places where the strings overlap, where overlap is defined using
some radius $r\sim\eta^{-1}$.  The corrections discussed above might
lead, at most, to a change of $r$ by a small numerical factor, but our
conclusions would not be affected.

The position of the string at time $t$ can be given by a function $\bx
(\sigma, t)$.  With the usual choice of parameterization (i.e., gauge)
the function $\bx$ satisfies
\blea
|\bx' (\sigma, t)| ^ 2 + |\xdot (\sigma, t)|^2 & = & 1\label{eqn:xpxd}\\
\bx' (\sigma, t)\cdot\dot\bx (\sigma, t) & = & 0
\elea
and the equation of motion is
\be
\bx'' (\sigma, t) =\ddot\bx (\sigma, t)
\ee
where $\bx'$ denotes differentiation with respect to $\sigma$ and
$\dot\bx$ denotes differentiation with respect to $t$.
The general solution is
\be\label{eqn:general}
\bx (\sigma, t) =\half (\ba (\sigma - t) +\bb (\sigma + t))
\ee
where $\ba$ and $\bb$ are arbitrary functions that satisfy $|\ba' | =
|\bb' | = 1$.  A cusp is a point at which $\ba' (\sigma -t) = -\bb'
(\sigma + t)$ and thus $\bx' = 0$ and $|\dot\bx | = 1$.

We will expand $\bx$, $\ba$ and $\bb$ in Taylor series around the
position of the cusp, which we will take as $\sigma = 0$ and $t = 0$.
By dimensional arguments, for a string with a typical length scale $L$
we expect the $n$th derivatives to be of order $L ^ {1-n}$, whereas,
as we will see, we expect $\sigma$ to be of order $\sqrt {rL}$, where
$r$ is the string radius.  Since $r$ is tiny (only a few orders of
magnitude above the Planck scale for a GUT-scale gauge string), while
$L$ is a cosmological size, we see that this expansion converges very
rapidly indeed.

\begin{figure}
\begin{center}
\leavevmode\epsfbox{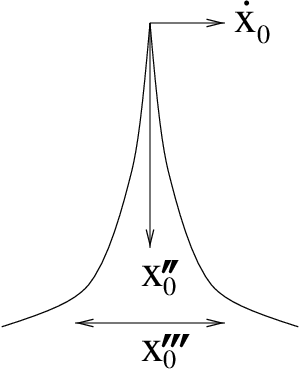}
\end{center}
\caption{The parameters of a cusp.  The tip moves at the speed of
light in the direction $\xdot_0$, the direction of the string near the
cusp is given by $\bx''_0$, and the spreading of the strings is in the
direction $\bx'''_0$.}
\label{fig:cusp}
\end{figure}

We will take the origin of coordinates at the cusp and expand $\ba$
and $\bb$ near the cusp to third order in $\sigma$,
\blea
\ba (\sigma) & = &\ba'_0\sigma +\ba''_0 {\sigma ^ 2\over 2} +\ba'''_0
{\sigma ^ 3\over 6}\\
\bb (\sigma) & = &\bb'_0\sigma +\bb''_0 {\sigma ^ 2\over 2} +\bb'''_0
{\sigma ^ 3\over 6}
\elea
where subscript $0$ denotes quantities at the cusp.
To have a cusp we require 
\be\label{eqn:d1}
\ba'_0 = -\bb'_0
\ee
from which $\xdot_0 =\bb'_0$ and $\bx'_0 = 0$.
Similarly, we can expand
\be\label{eqn:xexp}
\bx (\sigma) =\bx''_0 {\sigma ^ 2\over 2} +\bx'''_0 {\sigma ^ 3\over 6}\,.
\ee
These vectors are shown in Fig.\ \ref{fig:cusp}.

To maintain $|\ba' (\sigma) | = 1$ and $|\bb' (\sigma) | = 1$ we
require
\bml\label{eqn:orthogonal}\bea
\ba''\cdot\ba' & = & 0\label{eqn:210}\\
\ba'''\cdot\ba' & = & -|\ba'' | ^ 2\label{eqn:3122}
\elea
and similarly for $\bb$.
From Eqs.\ (\ref {eqn:d1}) and (\ref {eqn:orthogonal}) we see that
\blea
\bx''_0\cdot\dot\bx_0 & = & 0\label{eqn:o2}\\
\bx'''_0\cdot\dot\bx_0 & = & \half (|\ba''_ 0 | ^ 2-|\bb''_0 | ^ 2)\,.
\label{eqn:o3}
\elea

Some authors \cite{Branden87,Pijus89,Branden90,Branden93a,Branden93b} 
have considered cusps with $\bx'''_0\cdot\dot\bx_0 = 0$,
but in general this will not be the case.  We will see later that
$|\ba''_0 |/|\bb''_0 |$ is unaffected by Lorentz transformations, so
that a cusp which does not have $\bx'''_0\cdot\dot\bx_0 = 0$ cannot be
transformed into one which does.

We can align our coordinate system so that $\xdot_0$ lies along the
positive $x$-axis.  Then $\ba''_0$ and $\bb''_0$ lie in the $y$-$z$
plane.  Since the overall orientation in this plane is immaterial, the
degrees of freedom at the second derivative level are the magnitudes
$|\ba''_0 |$ and $|\bb''_0 |$ and the angle between $\ba''_0$ and
$\bb''_0$.  Similarly at the third derivative level, the component of
$\ba'''_0$ in the $x$ direction is determined by Eq.\ (\ref{eqn:3122}),
and similarly for $\bb'''_0$, so we have two degrees of freedom each
for $\ba'''_0$ and $\bb'''_0$.  The cusp is specified by seven
parameters in all.

\section {Lorentz transformations}
We now consider the effect of a Lorentz boost on the second- and
third-derivative parameters which characterize our cusp.  To determine
the transformation we proceed as follows.  The string world sheet is a
2-dimensional surface in 4-dimensional space.  The surface is timelike
except at isolated cusp points, where it is null.  A timelike
2-surface can be characterized at each point by two forward-directed
null vectors $A$ and $B$.  The length of these vectors is arbitrary.

In a particular coordinate system, we can express the position of the
string by a function $\bx (\sigma, t)$, which we expand as in
Eq.\ (\ref{eqn:general}).
The 4-vector $(0,\bx')$ represents motion along the string at fixed
$t$ and thus is tangent to the world sheet.  Similarly, the 4-vector
$(1,\xdot)$ represents motion into the future and in the direction of
motion of the string, so it also is tangent to the world sheet.  The
sums and differences of these vectors are null and can be taken as our
$A$ and $B$,
\blea
A^\mu & = & (1,\xdot - \bx') = (1, -\ba')\\
B ^\mu & = & (1,\xdot + \bx' ) = (1,\bb')\,.
\elea

If we now transform to a new coordinate system, the vectors $A$ and $B$
will still be tangent to the world sheet.  These vectors will still be
null but they will have new components which we will write
\blea
A ^\mutilde & = & (A ^ \ttilde, {\bf A})\\ B ^\mutilde & = & (B ^ \ttilde, {\bf
B})\,.
\elea
We will denote by $\atilde$ and $\btilde$ the functions $\ba$ and
$\bb$ in the new coordinate system.  We can determine $\atilde'$ and
$\btilde'$ by simply scaling $A ^\mutilde$ and $B ^\mutilde$,
\blea
\Atilde ^\mutilde &= & (1,-\atilde') = A/A ^ \ttilde\\
\Btilde ^\mutilde  &= & (1,\btilde') = B/B ^ \ttilde\,.
\elea

We will let our new coordinate system move with velocity
$-\bbeta$, so that a particle at rest in the original system is moving
with velocity $\bbeta$ with respect to the new coordinates.  The
Lorentz transformation then gives $A ^\ttilde =\gamma (1
-\ba'\cdot\bbeta)$ and $B ^\ttilde =\gamma (1 +\bb'\cdot\bbeta)$, where
$\gamma = 1/\sqrt {1-\beta ^ 2}$.  We will define
\blea
f_A & = & {1\over A ^\ttilde} = {1\over\gamma (1-\ba'\cdot\bbeta)}
\label{eqn:fAdef}\\
f_B & = & {1\over B ^\ttilde} = {1\over\gamma (1+\bb'\cdot\bbeta)}\,.
\elea

Now let $h$ be any function defined near the world sheet.  We can
differentiate in the direction of $A$ or $B$,
\bea
A ^\mu\partial_\mu h &=& {\partial h\over\partial t}\bigg|_\bx
-\ba'\cdot {\bf\nabla} h
={\partial h\over\partial t}\bigg|_\bx +\xdot\cdot {\bf\nabla}
h-\bx'\cdot {\bf\nabla} h \nonumber\\
&=& {\partial h\over\partial t}\bigg|_\sigma -{\partial
h\over\partial\sigma}\bigg|_t\label{eqn:dAdef}
\eea
and
\bea
B ^\mu\partial_\mu h &=& {\partial h\over\partial t}\bigg|_\bx
+\bb'\cdot {\bf\nabla} h
={\partial h\over\partial t}\bigg|_\bx +\xdot\cdot {\bf\nabla}
h+\bx'\cdot {\bf\nabla} h\nonumber\\
&=& {\partial h\over\partial t}\bigg|_\sigma +{\partial
h\over\partial\sigma}\bigg|_t\,.
\eea

If we make an arbitrary extension of $A$ and $B$ to a neighborhood of
the world sheet, we can define
\blea
B_2^\nu & = & B ^\mu\partial_\mu B ^\nu = (0,\dot\bb' +\bb'') = (0,2\bb'')\\
A_2^\nu & = & A ^\mu\partial_\mu A ^\nu = (0,-\dot\ba' +\ba'') = (0, 2\ba'')\,.
\elea
In the new reference frame we will have
$\Atilde_2 ^\mutilde = (0, 2\atilde'')$ where
\bea\label{eqn:A2}
\Atilde_2 &= &\Atilde ^\mutilde\partial_\mutilde\Atilde = f_A A
^\mu\partial_\mu (f_AA) = f_A ^ 2 A ^\mu\partial_\mu A + f_Af_{A, A} A
\nonumber\\
&= & f_A ^ 2A_2 + f_Af_{A, A} A
\eea
with $f_{A, A} = A ^\mu\partial_\mu f_A$, and similarly
$\Btilde_2 ^\mutilde = (0, 2\btilde'')$ with
\bea\label{eqn:B2}
\Btilde_2 &=&\Btilde ^\mutilde\partial_\mutilde\Btilde = f_B B
^\mu\partial_\mu (f_BB) = f_B ^ 2 B ^\mu\partial_\mu B + f_Bf_{B, B} B
\nonumber\\
&= & f_B ^ 2B_2 + f_Bf_{B, B} B
\eea
and $f_{B, B} = B ^\mu\partial_\mu f_B$.

We are not concerned with the actual directions of $\ba'$, $\ba''$,
and so on, but only their lengths and the angles between them.  We can
compute
\bea
4 |\atilde''| ^ 2 = g (\Atilde,\Atilde)
&=& f_A ^ 2f_{A, A} ^ 2g (A, A) + 2f_A ^
3f_{A, A} g (A, A_2)\nonumber\\
& & + f_ A ^ 4g (A_2,A_2)
\eea
where we have used a metric $g =\diag (-1,1,1,1)$.

Since $A$ is null, $g (A, A) = 0$.  We also have $g (A,A_2) =
2\ba'\cdot\ba''= 0$ from Eq.\ (\ref {eqn:210})
and so
\be
|\atilde''| = f_A ^ 2 |\ba''|\,.
\ee
Similarly, $|\btilde'' | = f_B ^ 2 |\bb'' |$.

At the cusp, $A = B$ and so we can let 
\be\label{eqn:fdef}
f =f_A = f_B\,.
\ee
At this point $g (A, B) = 0$ and $g (A_2,B) = g (B_2, A) = 0$,
so
\be
\atilde_0''\cdot\btilde_0'' = f ^ 4\ba''_0\cdot\bb''_0\,.
\ee
This means that the angle between $\ba''_0$ and $\bb''_0$ is unaffected by the
boost.  At the second derivative level, the effect of a Lorentz
transformation (up to rotation) is purely to rescale $\ba''_0$ and
$\bb''_0$ by the same factor $f ^ 2$.

We can compute the third derivatives using the same technique,
\blea
(0, -4\ba''')& = & A_3 ^\rho =A ^\mu\partial_\mu (A ^\nu\partial_\nu A ^\rho)\\
(0, 4\bb''') & = &B_3 ^\rho =B ^\mu\partial_\mu (B ^\nu\partial_\nu B ^\rho)
\elea
and
\blea \label{eqn:A3}
\Atilde_3  &= & f_A A ^\mu\partial_\mu (f_AA ^\nu\partial_\nu(f_AA))\nonumber\\
& =&f_A ^ 3 A_3 + 3f_A ^ 2f_{A, A} A_2 + (f_Af_{A, A} ^2 + f_A^2 f_{A, AA}) A\\
\Btilde_3  &= & f_B B ^\mu\partial_\mu (f_BB^\nu\partial_\nu(f_BB))\nonumber\\
&=& f_B ^ 3 B_3 + 3f_B ^ 2f_{B, B} B_2 + (f_Bf_{B, B} ^ 2 + f_B^2 f_{B, BB}) B
\elea
where $f_ {A, AA} = A ^\mu\partial_\mu (A ^\nu\partial_\nu f_A)$
and $f_ {B, BB} = B ^\mu\partial_\mu (B ^\nu\partial_\nu f_B)$.

The component of $\ba'''_0$ in the direction of $\xdot_0$ is fixed by
Eq.\ (\ref{eqn:3122}), so we are only interested in the components in
the directions of $\ba''_0$ and $\bb''_0$.
When we take the inner product of Eqs.\ (\ref{eqn:A3}) and 
(\ref{eqn:A2}), only the first and second terms of Eq.\
(\ref{eqn:A3}) will contribute as before:
\bea
-8\atilde'''\cdot\atilde''&=& g (\Atilde_3, \Atilde_2) = f_A ^ 5g
(A_3, A_2) + f_A ^ 4f_{A, A} g (A_3, A)\nonumber\\
&&\qquad\qquad\qquad + 3f_A^4f_{A, A} g (A_2, A_2)\nonumber\\
&=& -8f_A ^ 5\ba'''\cdot\ba''+ 4f_A ^ 4f_{A,
A}\ba'''\cdot\ba'\nonumber\\
&&\quad\quad+12f_A ^ 4f_{A, A} |\ba''| ^ 2\,.
\eea
Using Eq.\ (\ref{eqn:3122}) we can write this
\be\label{eqn:a3a1}
\atilde'''\cdot\atilde''= f_A ^ 5\ba'''\cdot\ba'' - f_A ^ 4f_{A, A}
|\ba''| ^ 2\,.
\ee
Now from Eqs.\ (\ref{eqn:fAdef}) and (\ref{eqn:dAdef}),
\bea
f_{A, A}  &= & {\partial f_A\over\partial t }\bigg|_\sigma -{\partial
f_A\over\partial\sigma}\bigg|_t = {\bbeta\over\gamma
(1-\ba'\cdot\bbeta) ^ 2} \cdot (\dot\ba'-\ba'')\\  &= & -f_A
{2\bbeta\cdot\ba''\over 1-\ba'\cdot\bbeta} = -2\gamma f_A ^
2\bbeta\cdot\ba''
\eea
and similarly
\bea
f_{B, B}  &= & {\partial f_B\over\partial t }\bigg|_\sigma +{\partial
f_B\over\partial\sigma}\bigg|_t = { -\bbeta\over\gamma
(1+\bb'\cdot\bbeta) ^ 2} \cdot (\dot\bb'+\bb'')\\  &= & -f_B
{2\bbeta\cdot\bb''\over 1+\bb'\cdot\bbeta} = -2\gamma f_B ^
2\bbeta\cdot\bb''\,.
\eea
Equation (\ref{eqn:a3a1}) then becomes
\be
\atilde'''\cdot\atilde''= f_A ^ 5 (\ba'''\cdot\ba''+ 2\gamma f_A
|\ba''| ^ 2\bbeta\cdot\ba'')\,.
\ee
In the same manner we can compute
\be
\btilde'''\cdot\btilde''= f_B ^ 5 (\bb'''\cdot\bb'' -2\gamma f_B |\bb''|
^ 2\bbeta\cdot\bb'')
\ee
and, at the cusp, using Eq.\ (\ref{eqn:fdef}),
\blea
\atilde_0'''\cdot\btilde''_0& = & f ^ 5 [\ba'''_0\cdot\bb''_0 \nonumber\\
& &\qquad+\gamma f (3
(\ba_0''\cdot\bb''_0)\bbeta\cdot\ba''_0-|\ba''_0| ^ 2\bbeta\cdot\bb''_0)]\\
\btilde'''_0\cdot\atilde''_0& = & f ^ 5 [\bb'''_0\cdot\ba''_0\nonumber\\
& &\qquad+\gamma f
(|\bb''_0| ^ 2\bbeta\cdot\ba''_0
-3(\ba_0''\cdot\bb''_0)\bbeta\cdot\bb''_0) ]\,.
\elea

Now we consider two types of Lorentz transformation: a longitudinal
boost in which $\bbeta$ is parallel to $\xdot_0$, and a boost with a
transverse component but which has $f = 1$.  For the longitudinal
boost, the directions of $\xdot_0$, $\ba''_0$ and $\bb''_0$ are
unaffected.  For the $f = 1$ boost we will rotate the system after the
boost so that these vectors are returned to their original directions.

For a longitudinal boost with $\bbeta =\beta\xdot_0$, the effect is to
rescale the various parameters,
\bml\label{eqn:longscale}\bea
\atilde_0''& = & f ^ 2\ba_0''\\ \btilde_0''& = & f ^ 2\bb _0''\\
\mbox{$\atilde'''_0$} ^ {(\perp)} & = & f ^ 3{\ba'''_0} ^ {(\perp)}\\
\mbox{$\btilde'''_0$} ^ {(\perp)} & = & f ^ 3{\bb'''_0} ^ {(\perp)}\\
\mbox{$\atilde'''_0$} ^ {(\parallel)} & = & f ^ 4{\ba'''_0} ^ {(\parallel)}\\
\mbox{$\btilde'''_0$} ^ {(\parallel)} & = & f ^ 4{\bb'''_0} ^ {(\parallel)}
\elea
where ${(\perp)}$ denotes components perpendicular to $\xdot_0$ and
${(\parallel)}$ denotes parallel components, and
\be
f = {1\over\gamma (1 +\beta)} = {\sqrt{1-\beta ^ 2}\over 1 +\beta}
=\sqrt{1-\beta\over 1 +\beta}\,.
\ee
We see that $1/f$ is just the Doppler shift factor for radiation
moving in the $\xdot_0$ direction.

For an $f = 1$ transformation, $\ba'_0$, $\bb'_0$,
${\ba''_0} ^ {(\parallel)}$, and ${\bb''_0} ^ {(\parallel)}$ are
unchanged, and
\blea
\atilde'''_0\cdot\atilde''_0&=&  \ba'''_0\cdot\ba''_0+ 2\gamma
|\ba''_0| ^ 2\bbeta\cdot\ba''_0\\
\atilde_0'''\cdot\btilde''_0& = &\ba'''_0\cdot\bb''_0 +\gamma (3
(\ba_0''\cdot\bb''_0)\bbeta\cdot\ba''_0-|\ba''_0| ^ 2\bbeta\cdot\bb''_0)\\
\btilde'''_0\cdot\atilde''_0& = & \bb'''_0\cdot\ba''_0+\gamma 
(|\bb''_0| ^ 2\bbeta\cdot\ba''_0 -
3 (\ba_0''\cdot\bb''_0)\bbeta\cdot\bb''_0)\\
\btilde'''_0\cdot\btilde''_0&=& \bb'''_0\cdot\bb''_0 -2\gamma |\bb''_0| ^ 2\bbeta\cdot\bb''_0\,.
\elea

There are two parameters which specify the $f = 1$ transformation.
For example, we can write
\be
\bbeta =\betapar\xdot_0 +\betaperp
\ee
where $\betaperp$ is perpendicular to $\xdot_0$.  To make $f = 1$ we
demand that $\gamma (1-\betapar) = 1$ or $\sqrt {1-\beta ^ 2} =
1-\betapar$, so $|\betaperp | ^ 2 = 2\betapar -2\betapar ^ 2$.  We
can specify $\betapar$ and also the direction of $\betaperp$ in the
plane perpendicular to $\xdot_0$.

By the use of these two degrees of freedom, we can impose two
constraints on $\ba'''_0$ and $\bb'''_0$.  For example, we can require
that $\bx'''_0$ be parallel to $\xdot_0$, as follows.
We seek an $f = 1$ transformation such that
\blea
0 & = &\xtilde'''_0\cdot\atilde''_0 =\bx'''_0\cdot\ba''_0 +
\gamma\bbeta\cdot \bigg(\left(|\ba''_0 | ^ 2 +\half |\bb''_0 | ^ 2\right)\ba''_0\nonumber\\
&&\qquad\qquad\qquad\qquad\qquad\qquad-{3\over 2}(\ba''_0\cdot\bb''_0)\bb''_0\bigg)\\
0 & = &\xtilde'''_0\cdot\btilde''_0
=\bx'''_0\cdot\bb''_0 + \gamma\bbeta\cdot \bigg({3\over 2}
(\ba''_0\cdot\bb''_0)\ba''_0 \nonumber\\
&&\qquad\qquad\qquad\qquad\qquad-\left(|\bb''_0 | ^ 2+\half
|\ba''_0 | ^ 2\right)\bb''_0\bigg)\,.
\elea

These equations can be written in the form
\blea
\gamma\bbeta\cdot\balpha_1 & = & c_1\\ \gamma\bbeta\cdot\balpha_2 & =
& c_2
\elea
where $\balpha_1$ and $\balpha_2$ are vectors in the plane of
$\ba''_0$ and $\bb''_0$, and $c_1$ and $c_2$ are constants.  Assuming
that the vectors $\ba''_0$ and $\bb''_0$ are linearly independent,
$\balpha_1$ and $\balpha_2$ will be also, since one can show that the
transformation matrix is not singular.  Thus we can choose a
direction $\betahatperp$ such that
\be
{\betahatperp\cdot\balpha_1\over c_1 } =
{\betahatperp\cdot\balpha_2\over c_2} \equiv c> 0\,.
\ee
Then we must find a value for $\betapar$ such that
\be
\gamma |\betaperp | = 1/c\,.
\ee
Since $f = 1$, $\gamma = 1/(1 -\betapar)$ and
\be
\gamma |\betaperp | = {|\betaperp|\over 1-\betapar} =\sqrt
{2\betapar\over 1-\betapar}
\ee
which can take any value by appropriate choice of $\betapar$.  Thus by
appropriate choice of parameters we can make $\bx'''_0\cdot\ba''_0
=\bx'''_0\cdot\bb''_0 = 0$ so that only the component of $\bx'''_0$
parallel to $\xdot_0$ survives.

If this cusp is part of a string with a typical length scale $L$, then
we expect $\ba''_0,\bb''_0\sim L ^ {-1}$ and $\bx'''_0\sim L ^ {-2}$.
Thus $|\balpha_1 |, |\balpha_2 |\sim L ^ {-3}$ and $c_1, c_2\sim L ^
{-3}$, so $\gamma |\beta |$ needs only to be of order 1.  We do not
need a huge boost to go to the frame where $\bx'''_0$ is parallel to
$\xdot_0$.

Once in this canonical frame, we can make longitudinal boosts to scale
$|\ba''_0 |$ and $|\bb''_0 |$.  Thus, up to boosts, the cusp is given
by four parameters: the relative magnitude of $\ba''_0$ and $\bb''_0$,
the angle between them, and two parameters giving $\xdot_0''$ in the
plane perpendicular to $\xdot_0$.

From these transformations we can determine the way in which the
amount of radiation produced by a cusp scales with the typical length
scale $L$.  We will imagine that we know nothing about the process by
which this radiation is produced, except for the following assumptions:
\begin{itemize}
\item The radiation is strongly relativistic.
\item The radiation is strongly beamed in the direction of $\xdot_0$.
\item The amount of radiation depends only on $\bx''_0$, $\xdot'_0$,
and $\bx'''_0$.
\end{itemize}

Consider a cusp with length scale $L$ and suppose that it emits an
amount of radiation $E$.  Now suppose first that we rescale this cusp
by a factor $s$.  We will have
\bml\label{eqn:scale}\bea
\bx''_0,\xdot'_0 &\sim & s ^ {-1}\\ \bx'''_0
&\sim & s ^ {-2}\,.
\elea
Now suppose instead that we transform the cusp by a longitudinal boost
with $f = s ^ {-1/2}$.  From Eqs.\ (\ref{eqn:longscale}) the parameters
scale as 
\blea\label{eqn:boostscale} 
\bx''_0,\xdot'_0 &\sim & f^2\sim s ^ {-1}\\
\bx'''_0 &\sim & f^4\sim s ^ {-2}
\elea
just as in Eqs.\ (\ref{eqn:scale}).  
Since the radiation is strongly relativistic and strongly
beamed in the forward direction, it will transform under the boost as
\be
E\longrightarrow\sqrt{1 +\beta\over 1-\beta} E = {E\over f} = s ^
{1/2} E \,.
\ee
Since the relevant parameters in the rescaled string are the same as
in the boosted string the rescaled string will also release energy $s
^ {1/2} E$.


\section {Overlap Calculation}
Outside the string core, the fields fall off rapidly to their vacuum
values.  Therefore, the portion of string near the cusp which can be
released as radiation can be at most the part where the string cores
overlap.  We will let $r$ denote the radius of the core at which this
annihilation becomes possible.  It is not clear precisely what this
value should be, but it is approximately the length corresponding to
the energy scale of the string, $r\sim 10 ^ {-30}\text{cm}$ for a GUT-scale
gauge string.

We will use the canonical reference frame discussed earlier, in which
$\xdot_{0}$ is parallel to $\bx'''_0$.\footnote{If we start with a
cusp in which $\bx'''_0$ is perpendicular to $\xdot_0$, in the
canonical frame we have $\bx'''_0 = 0$.  Thus to third order in $\sigma$
the two parts of the string lie on top of each other.  Only at fifth
order are they separate, so the radiation emitted will be much larger
than we calculate here.  If one analyzes this
cusp in the original frame, one finds that, at times either before or
after the time of the cusp, the string nearly crosses itself.  These
near crossings give rise to the large energy emission.
Since the typical cusp does not have the parameters chosen to produce
this exact alignment, this large amount of radiation occurs in only a
vanishingly small fraction of cases.}  We show in the appendix that,
in this frame, the length of the string overlap is maximized at the
time of the cusp ($t=0$).  At the order of approximation we will use,
the cusp is symmetrical, so we will consider only the overlap between
the point of the string at $\sigma$ and its symmetrical point at
$-\sigma$.

	Using the Taylor expansion, Eq.\ (\ref{eqn:xexp}), of the
string around the position of the cusp, the distance
between two points on the two different branches of the string is
given by
\be
|\bx (\sigma,0) - \bx (-\sigma,0)| =|\bx'''_0| {\sigma ^ 3\over 3} +
\cdots\cdot
\ee

Using Eq.\ (\ref{eqn:xpxd}) we can write the lowest order
correction to the velocity at a particular value of the parameter
$\sigma$,
\be
|\xdot (\sigma,0)|^2 = 1 - |\bx' (\sigma,0)|^2\,.
\ee
We can now use  Eq.\ (\ref{eqn:xexp}) to expand $\bx'$,
\be
\bx'(\sigma,0)= \bx''_0 \sigma + \bx'''_0 {\sigma^2\over 2} + \cdots
\ee
so at the lowest order we have
\be
|\xdot (\sigma,0)|^2 \simeq 1 - \sigma ^ 2 |\bx''_0 |^2\,.
\ee
The gamma factor at this order in $\sigma$ is
\be\label{eqn:gamma}
\gamma = {1\over \sqrt{ 1 - \xdot ^2}} \simeq {1\over |\bx''_0| \sigma}\,.
\ee

Due to Lorentz contraction of the core of the string in the 
direction of motion, the cross section of the string in 
this frame is not a circle of radius $r$, but an ellipse with
semimajor axis $r$ and semiminor axis $r/\gamma$ in the
direction of $\xdot$.

\begin{figure}
\begin{center}
\leavevmode\epsfbox{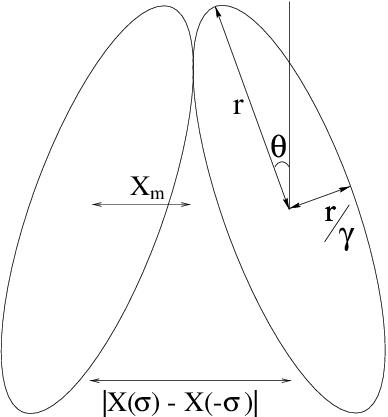}
\end{center}
\caption{A cross-section through the strings at the point of contact.} 
\label{fig:ellipses}
\end{figure}

Now the overlap calculation reduces to the computation of the value of
$\sigma$ at which these two ellipses touch one another.  As before we
will only take into account effects of order $\sigma$.  This allows us
to ignore the component of the velocity in the direction parallel to
$\bx''_0$.  The component of $\xdot$ in the plane perpendicular to
$\bx''_0$ is
\be 
\xdot_\perp = \xdot -  {(\xdot \cdot \bx''_0) \bx''_0 \over {|\bx''_0|^2}}\,.
\ee
Since $\xdot_0 \cdot \bx''_0 = 0$ we see that
\be 
\xdot_\perp = \xdot_0 + \sigma \bx_3\,,
\ee
where $\bx_3$ is the component of $\xdot'_0$ perpendicular to $\bx''_0$,
\be 
\bx_3 = \xdot'_0 - \left[ {(\xdot'_0 \cdot  \bx''_0) \over { |\bx''_0|^2}}\right] \bx''_0 \,.
\ee
At this order of approximation the angle between $\xdot_\perp$ and $\xdot_0$ is
\be
\theta \simeq {|\bx_3|\over |\xdot_\perp|} \sigma \,.
\ee
The cross section of the string in the plane perpendicular to
$\bx''_0$ can be approximated by two ellipses whose centers are
separated a distance $|\bx (\sigma) - \bx (-\sigma)|$ and whose major
axes are tilted towards each other, each by a small angle $\theta$.
See Fig.\ \ref{fig:ellipses}.

It can be shown that if an ellipse whose major axis is along the $y$
direction is tilted by a positive angle $\alpha$, the maximum value of $x$ that
the new rotated ellipse reaches is
\be
x_m = \sqrt { d_1^2 \sin^2 \alpha +  d_2^2 \cos^2 \alpha}
\ee
where $d_1$ and $d_2$ are the lengths of the semimajor and semiminor
axes respectively.
In our case,
\bea\label{eqn:xmtg}
x_m &=& \sqrt {r^2 \sin^2\theta +  {r^2\over \gamma^2} \cos^2 \theta}
\nonumber\\
&\simeq& r \sqrt{ \theta^2 + {1\over \gamma^2}} \simeq r \sigma \sqrt{ |\bx_3|^2 + |\bx''_0|^2}\,.
\eea

The maximum value of $\sigma$ at
which the two branches of the string touch is given by
\be
{\sigma_c^3 |\bx'''_0| \over 3} = 2 x_m
\ee
so
\bea
\sigma_c  &=& \left( { 6 r \sqrt{ |\bx_3|^2 + |\bx''_0|^2}\over {|\bx'''_0|}}\right)^{1/2}\nonumber\\
&=& \left( { 6r\over |\bx'''_0|}  \sqrt{ |\xdot'_0|^2 - 
 {(\xdot'_0 \cdot  \bx''_0)^2 \over { |\bx''_0|^2}}  +
|\bx''_0|^2}\right)^{1/2} \,.
\eea
Since $\bx'''_0$ is parallel to $\xdot_0$,  we can use Eq.\ (\ref{eqn:o3}),
\be
|\bx'''_0|= |\bx'''_0 \cdot \xdot_0|
={1\over 2}\left||\ba''_0|^2 -|\bb''_0|^2\right| = 2 
|\xdot'_0 \cdot \bx''_0|\,,
\ee
so we can write the expression for $\sigma_c$ as
\be
\sigma_c = \left(  3 r \sqrt{{{|\bx''_0|^2 + |\xdot'_0|^2}\over {|\xdot'_0 \cdot \bx''_0|^2}} 
- {1\over {|\bx''_0|^2}}}\right)^{1/2}
\ee
or in terms of the derivatives of the vectors $\ba$ and $\bb$ as

\be
\sigma_c = \left(  6 r \sqrt{  { 2 (|\ba''_0|^2+|\bb''_0|^2)\over 
 (|\bb''_0|^2 - |\ba''_0|^2)^2}- {1\over{(\ba''_0+\bb''_0)^2}}}\right)^{1/2}\,.
\ee
As discussed earlier, we expect the $n$th 
derivatives to be of the order of $L^ {1-n}$, so we finally get that
\be\label{eqn:sc}
\sigma_c \sim \sqrt { r L} + O(r)\,.
\ee
The segment of string whose energy could be radiated runs from
$-\sigma_c$ to $\sigma_c$, so the maximum energy release is
\be
E = 2 \mu \sigma_c \sim \mu  \sqrt { r L}\,,
\ee
where $\mu$ is the energy per unit length of the string.

Using Eqs.\ (\ref{eqn:gamma}) and (\ref{eqn:sc}) we can compute the
Lorentz boost of the string at the point where the overlap begins,
\be
\gamma_c\sim {1\over |\bx''_0 |\sigma_c}\sim\sqrt {L\over r}\,.
\ee
For a cosmological string, $L\gg r$ and so $\gamma_c\gg 1$.  In fact this
boost will normally be so large that the emitted particles will have
energies much larger than the Planck scale.

We imagine that the radiation consists of ``X bosons'' with rest mass
$m_X\sim\sqrt {\mu}$, so the number of particles emitted is
\be
N_X\sim {E\over\gamma m_X}\sim r\sqrt {\mu}\sim 1
\ee
independent of the cosmological length scale $L$.  The physical
size of the region from which these particles are emitted is
\be
l\sim |\bx''_0 |\sigma_c ^ 2\sim r
\ee
also independent of $L$.  Thus the radiation from a cusp consists of a
small number of particles with very large boosts emitted from a region
whose size is similar to the string thickness.


\section {Discussion}

We have analyzed the transformation properties of cosmic string cusps
under Lorentz boosts.  A generic cusp can be brought into a canonical
form in which the motion of the cusp tip lies in the plane of the
cusp.  In general a boost of large magnitude is not needed to go to
this frame.  Once in such a frame, the cusp can be further boosted to
scale one parameter, e.g., $\bx''$, to any desired value.  All the
second-derivative parameters scale together under this transformation.
Thus the difference between a cusp that one would expect to find in a
large loop and one that one would expect in a small loop is essentially a
matter of boosting.

The maximum amount of radiation which can be emitted from a cusp is
given by the energy stored in the parts of the string whose cores
overlap.  Taking into account the Lorentz contraction of the core due
to the rapid motion of the string near the cusp, we have found that
the emitted energy is at most of order $\mu\sqrt {rL}$.  Previous
analyses \cite{Branden87,Pijus89,Branden90,Branden93a,Branden93b}
have used the result $\mu r ^ {1/3} L ^ {2/3}$; the present
result differs by a factor $(r/L) ^ {1/6}$.  Since $r$ is of
microphysical size, while $L$ is cosmological, the energy emitted is
reduced by many orders of magnitude by this effect.  Since even
neglecting Lorentz contraction the radiation from cusps would at most
be barely detectable \cite{Pijus89,Branden90,Branden93a,Branden93b},
 this effect prevents any such observation.

We would like to thank Xavier Siemens and Alex Vilenkin for helpful
conversations.  This work was supported in part by funding provided by
the National Science Foundation.  J. J. B. P. is supported in part by the
Fundaci\'on Pedro Barrie de la Maza.

\appendix
\section {Time variation}

The computation of the value of $\sigma$ at which the two branches of
the string overlap involves three different effects which
have different time-dependence, so there is the possibility that
portions of the string might overlap at times $t > 0$ or $t<0$ that are
separate at $t = 0$.  In order to clarify this point, we redo the
calculation, keeping the dependence on time up to second order.  Taking 
into account that in this frame $\bx'''_0 \cdot \xdot'_0 =0$, the
distance between the string centers is now
\be
\label{eqn:dis}
|\bx (\sigma,t) - \bx (-\sigma,t)|^2 \simeq {\sigma^6\over9} |\bx'''_0|^2
 +   4 t^2 \sigma^2 |\xdot'_0|^2
 + \cdots\,,
\ee
the Lorentz factor is
\be
\gamma = {1\over \sqrt{ 1 - \xdot ^2}} = {1\over |\bx'|}
\ee
so
\be
\label{eqn:gam}
{{1}\over{\gamma^2}} \simeq |\bx''_0|^2 \sigma^2 
+2 (\bx''_0 \cdot \xdot'_0) \sigma t +  |\xdot'_0|^2 t^2  \cdots
\ee
and the angle of the ellipses is
\bea
\label{eqn:thet}
\theta^2  \simeq |\xdot_\perp - \xdot'_0|^2 \simeq  && \left(|\xdot'_0|^2 - 
 {|\xdot'_0 \cdot  \bx''_0|^2 \over { |\bx''_0|^2}}\right) \sigma^2 
\nonumber\\
& & + O(\sigma t^2) + \cdots\,.
\eea
Using Eq.\ (\ref{eqn:xmtg}) we can compute the radius of the string
core which would make the two branches of the string touch at $\sigma$,
\be
r(\sigma, t)^2 = {{|\bx (\sigma,t) - \bx (-\sigma,t)|^2}\over
 {\theta^2 + 1/\gamma^2}}
\ee
which can be written in terms of the dependence on $\sigma$ and $t$ as
\be\label{eqn:orders}
r(\sigma, t)^2 = {{O({\sigma ^6}) + O(t^2 \sigma^2)}\over { O(\sigma^2)
 + O(\sigma t) + O( t^2)}}\,.
\ee
We now expand this function around $t=0$,
\be
r(\sigma, t)^2 = r(\sigma, 0)^2 + \left({dr^2\over dt}\right)_0 t + 
 \left({d^2r^2\over dt^2}\right)_0 {t^2\over 2} +\cdots
\ee
with
\bml \label{eqn:rexp}\bea
 r(\sigma, 0)^2  &=& O(\sigma^4)\\
\left({dr^2\over dt} \right)_0 &=& O(\sigma^3)\\
\left({d^2r^2\over dt^2} \right)_0 &=& O(1)\,.
\elea
It can also be shown from the Eqs.\ (\ref{eqn:dis}), (\ref{eqn:gam})
and (\ref{eqn:thet}) that $\left({d^2r^2/ dt^2} \right)_0$ 
is positive so we can obtain from Eqs.\ (\ref{eqn:rexp})
the time at which  $r (\sigma, t)$ is minimum, namely 
\be 
t_{min} = O(\sigma^3)\,.
\ee
If we use this order of $\sigma$ in Eq.\ (\ref{eqn:orders}),
we see that the minimum value of $r^2$ is, at our order of
approximation, the same as $r(\sigma, 0)^2$, so we can compute the
value of $\sigma_c$ using $r (\sigma_c, 0) = r$, as above.

\bibliography{cosmic-string}
\bibliographystyle{prsty}

\end{document}